# Spectral fluctuation properties of spherical nuclei

M. A. Jafarizadeh[a,b1], N. Fouladi[c], H. Sabri[c2]

[a]Department of Theoretical Physics and Astrophysics, University of Tabriz, Tabriz 51664, Iran.

[b] Research Institute for Fundamental Sciences, Tabriz 51664, Iran.

[c] Department of Nuclear Physics, University of Tabriz, Tabriz 51664, Iran.

[1] E-mail:jafarizadeh@tabrizu.ac.ir
[2] E-mail:h-sabri@tabrizu.ac.ir

## Abstract

The spectral fluctuation properties of spherical nuclei are considered by use of NNSD statistic. With employing a generalized Brody distribution included Poisson, GOE and GUE limits and also MLE technique, the chaoticity parameters are estimated for sequences prepared by all the available empirical data. The ML-based estimated values and also KLD measures propose a non regular dynamic. Also, spherical odd-mass nuclei in the $50 < A \leq 100$ mass region, exhibit a slight deviation to the GUE spectral statistics rather than the GOE.

**Keywords:** Spectral statistics- spherical nuclei - Maximum Likelihood Estimation (MLE) – Gaussian Unitary Ensemble, Kullback-Leibler Divergence (KLD)

**PACS:** 24.60.-k, 23.20.-g, 02.50.Tt

## Introduction

The Random Matrix Theory (RMT) is known as a main tool in describing the statistical distribution of the energy-eigenvalues for the quantum counterpart of a classical chaotic system [1-2]. It has been supposed that eigenvalues belonging to different irreducible representations of the symmetry group are statistically independent and obey Gaussian Orthogonal Ensemble (GOE) or Gaussian Unitary Ensemble (GUE) spectral statistics [3-5]. GUE statistics usually is expected for spectra of non-time-reversal invariant systems, while time-reversal invariant (TRI) systems with integral spin or rotational invariance should explain level-distributions according to GOE. Recently it has been exposed [5-10], time-reversal invariant systems with discrete symmetries may display, in certain irreducible subspaces, the spectral statistics corresponding to the Gaussian-unitary ensemble (GUE) rather than to the expected orthogonal one (GOE) [5]. Kramers-type degeneracy is supposed in such situations while similar degeneracy is expected for spherical odd-mass nuclei. Different statistics such as Nearest Neighbor Spacing Distribution (NNSD) [11-12] and etc [13-15] have been proposed to exhibit the statistical situation of systems in related to regular (Poisson limit) and chaotic (GOE or GUE) limits while NNSD is the observable most commonly used to analyze the short-range fluctuation properties in the nuclear spectra.

Since, the NNSD involves complete (few or no missing levels) and pure (few or no unknown spin-parities) level scheme, to achieve a significant statistical analysis [3-4], we in need to combine different level schemes to prepare sequences. In the present study, we consider the statistical properties of spherical nuclei [16-20] classified in different mass regions. To estimate with more precision, we generalized Brody distribution (with more parameters) which considered all Poisson, GOE and GUE limits [21]. With help of the MLE technique [22], the essential relations have been evaluated[19] while explore very exact results with low uncertainty. Sequences prepare by only $2^{+}$levels for even mass and by $\frac{1}{2}^{+}$ levels for odd-mass nuclei (for their relative abundance) with all available empirical data [23-26] of nuclei in which the spin-parity $J^{\pi}$ assignment of at least five consecutive levels are definite. The ML-based estimated values for chaoticity parameters, propose a deviation to non regular dynamic for all sequences while as have predicted by Bae et al [27], heavier nucleus explore less chaoticity in compare to light ones. Also, spherical odd-mass nuclei propose more chaotic statistics in compare to even-mass ones in all categories. The KLD measure consider the distances of estimated distribution to Poisson, GOE and GUE limits, suggest a trivial deviation to GUE spectral statistics for spherical odd-mass nuclei in the $50 < A \leq 100$ mass region.

This paper is organized as follows: section 2 briefly summarizes the theoretical aspects of spherical nucleus, section 3 dealt with reviewing a statistical approach contained generalized Brody distribution and MLE method and finally, section 4 includes the numerical results obtained by applying the MLE to different sequences. Section 5 is devoted to summarize and some conclusion based on the results given in section 4.

## 2. Spherical nuclei

In the geometrical collective model first suggested by Bohr and Mottelson [16], nucleus modeled as a charged liquid drop. This model exhibits the moving of nuclear surface by an expansion in spherical harmonic with time-dependent shape parameters as coefficients [18]:

$$R(\theta,\varphi,t) = R_{av}\left[1+\sum_{l=0}^{\infty}\sum_{\mu=-l}^{l}\alpha_{l\mu}(t)Y_{l\mu}(\theta,\varphi)\right], \tag{2.1}$$

Where $R(\theta,\varphi,t)$ represent the nuclear radius in the direction $(\theta,\varphi)$at time $t$ and $R_{av}$ is the radius of the spherical nucleus. Different $l-$ values denote a translation or deformation seems at the collective excitation of nucleus. $l$ and $\mu$ establish the surface coordinate as function of $\theta$ and $\varphi$ respectively. For axially symmetric nuclei, the nuclear radius rewrite as:

$$R(\theta,\varphi) = R_{av}[1+\beta_2 Y_{20}(\theta,\varphi)], \quad (2.2)$$

The quadrupole deformation parameter $\beta_2(=\alpha_{20})$, is according to the axes of the spheroid by

$$\beta_2 = \frac{4}{3}\sqrt{\frac{\pi}{5}}\frac{\Delta R}{R_{av}}$$

In which the average radius, $R_{av} = R_0 A^{1/3}$, and $\Delta R$ is the difference between the semimajor and semiminor axes where the larger value of $\beta_2$ explore more deformation of nuclei [18]. We qualify nuclei as deformed according to the liquid drop model calculation by P.Moller et al. [26], therefore, spherical nuclei for which the deformation parameter $\beta_2$ is equal to zero classified in different mass region and also even- and odd-mass groups. In order to prepare sequences by different nucleus with the available empirical data taken from [23-26], we have followed the same method given in Ref.[3]. Namely, we consider nuclei in which the spin-parity $J^{\pi}$ assignments of at least five consecutive levels are definite. In cases where the spin-parity assignments are uncertain and where the most probable value appeared in brackets, we admit this value. We terminate the sequence in each nucleus when we reach at a level with unassigned $J^{\pi}$. We focus on $2^{+}$levels for even mass nuclei and $\frac{1}{2}^{+}$ levels for odd-mass ones for their relative abundance in specified nuclei. In this approach, we achieved 90 spherical nuclei (65 even-mass in addition to 25 odd-mass nuclei) as presented in Table1.

## 3. Nearest Neighbor Spacing Distribution

The fluctuation properties of nuclear spectra have been considered by different statistics. All of them (such as Nearest Neighbor Spacing Distribution (NNSD) [11-12], the Dyson-Mehta $\Delta_3$statistic [13-15] and etc) have been carried with comparison of fluctuation properties of selected spectrum with theoretical predictions of Random Matrix Theory (RMT), integrable (ordered) systems or interpolation between these two chaotic and regular limits. In NNSD method, level spacing of nuclear spectra have been prepared with unfolding processes to compare with theoretical accounts. The distribution $P(s)$ is the best spectral statistic to analyze shorter series of energy levels and the intermediate regions between order and chaos. To unfold our spectrum, we must use some levels with same symmetry [3].This requirement means to use levels with same total quantum number (J) and same parity. Then we first include the number of the levels below $E$ and write it as [21]

$$N(E) = N_{ave}(E) + N_{fluc}(E)$$

Then with taking a smooth polynomial function of degree 6 to fit the staircase function, we fix $N_{ave}(E)$. Therefore, the unfolded spectrum with the mapping $E_i \to \epsilon_i$

$$\epsilon_i = N_{ave}(E_i)$$

is prepared. The nearest-neighbor level spacing is defined as $s_i(\equiv \epsilon_{i+1} - \epsilon_i)$ which unfolded sequence $\{s_i\}$ is clearly dimensionless and has a constant average spacing of 1,then distribution $P(s)$ will be as $P(s)ds$ that is the probability for the $s_i$ to lie within the infinitesimal interval [s,s+ds]. It has been shown that the nearest-neighbor spacing distribution $P(s)$ measures the level repulsion. It has been assumed that eigenvalues belonging to different irreducible representations of the symmetry group are statistically independent and obey GOE [1]

$$P(s) = \frac{1}{2}\pi s e^{-\frac{\pi s^2}{4}} , \tag{3.1}$$

Or GUE spectral statistics

$$P(s) = \frac{32}{\pi^2} s^2 e^{-\frac{4s^2}{\pi}} . \tag{3.2}$$

Depending on whether TRI holds or not. On the other hand, in non-interacting systems (where a number of vanishing H-matrix elements appear because of the presence of certain symmetries such as isospin symmetry, the NNSD is well approximated by Poisson distribution [1]

$$P(s) = e^{-s} , \tag{3.3}$$

Spectral statistics of different considered systems exhibit interpolation between these limits while prove the theoretical predictions about mixture of regular and chaotic dynamics for low-lying energy levels of excited nuclei [3-4]. In order to quantify the chaoticity of $P(s)$ in terms of a parameter, it can compare for example with the Brody distribution [28] and etc [29-30], which are adequate for description of intermediate situations between order and chaos. Consequently, their parameters explain the chaoticity degrees of systems. To investigate all limits (Poisson, GUE and GOE), and also improve the precisions of estimated values, we generalized [21] the Brody distribution which considered Poisson (regular), GOE and GUE (chaotic) statistics and also estimated with more precision. We extended both (3.1) to (3.3) relations by means of ansatz

$$P(s) = b(1+q)(\alpha s^q + \beta s^{q+1}) e^{-bs^{q+1}}, \tag{3.4}$$

Where the constants obtained as

$$\alpha = 1 - \frac{\left(\frac{\Gamma\left[\frac{q+2}{q+1}\right]}{b^{\frac{1}{1+q}}}\right)^2 - \frac{\Gamma\left[\frac{q+2}{q+1}\right]}{b^{\frac{1}{1+q}}}}{\left(\frac{\Gamma\left[\frac{q+2}{q+1}\right]}{b^{\frac{1}{1+q}}}\right)^2 - \frac{\Gamma\left[\frac{q+3}{q+1}\right]}{b^{\frac{2}{1+q}}}} \quad , \quad \beta = \frac{\left(\frac{\Gamma\left[\frac{q+2}{q+1}\right]}{b^{\frac{1}{1+q}}}\right) - 1}{\left(\frac{\Gamma\left[\frac{q+2}{q+1}\right]}{b^{\frac{1}{1+q}}}\right)^2 - \frac{\Gamma\left[\frac{q+3}{q+1}\right]}{b^{\frac{2}{1+q}}}} \tag{3.5}$$

Which interpolate between Poisson ($q = 0 \& \beta = 0$), GOE ($q = 1 \& \beta = 0$) and GUE ($q = 1 \& \alpha = 0$) limits. To estimate with more precision, Maximum Likelihood Estimation (MLE) method [22] were employed while yield very exact results with low uncertainty in compare to other estimation methods particularly Least Square (LS) ones. The ML-based estimation method has been described in detail in Ref [21-22]. Here, we briefly outline the basic ansatz and summarize the results. The parameters b and q would estimate by high precision via solving related equations (relations (I-10,11) of Ref.[21]) by Newton-Raphson iteration method. Since, the exploration of the majority of short sequences yields an overestimation about the degree of chaoticity measured by the "$q, b$", therefore, we don't concentrate only on the implicit value of "q" and examine a comparison between the amounts of "$q$ and $b$" in the same mass region for different categories. Also, to compare the distance of estimated distribution with both GOE and GUE in some sequence exhibit a deviation to less regular dynamics ($q \rightarrow 1$), we evaluated Kullback-Leibler Divergence (KLD) measure as: [$P(i)$ represent our ML-based estimated distribution and GOE or GUE can be regard as $Q(i)$ where summation is carried on all $s_i$ of sequences][21]:

$$D_{KL}(P\|Q) = \sum_i P(i) \log \frac{P(i)}{Q(i)} \tag{3.6}$$

In which it would display closer distances between two distributions if $D_{KL}(P\|Q) \rightarrow 0$.

## 4. Numerical results

In the present study, we look the statistical properties of spherical nuclei. With regard to complete theoretical studies [16-18] and also experimental evidences [18-20], we tend to classify nuclei in different mass regions. Since we consider only sequences with at least 25 spacings are included, therefore, the 6 considered sequences are unfolded and analyzed with the help of MLE method. The ML–based estimated values for the parameters of generalized Brody distribution yield as the converging values of iterations by related equations [(I-10,11) of Ref.[21]], where as an initial values, the LS-based estimated values included. A comparison of the spacing distribution carried by the values of $(q, b)$ for these sequences are presented in Table1.Also, the NNSD histograms for even- and odd- mass nuclei represented in Figure (1 and 2) respectively.

| $Mass\ region$ | $q$ | $b$ | $KLD_{GOE}$ | $KLD_{GUE}$ |
|---|---|---|---|---|
| $A \leq 50$ $(even\ mass)$ | $0.72 \pm 0.04$ | $1.16 \pm 0.06$ | 1.174 | 3.211 |
| $A \leq 50$ $(odd\ mass)$ | $0.80 \pm 0.07$ | $1.10 \pm 0.07$ | 0.837 | 2.551 |
| $50 < A \leq 100$ $(even\ mass)$ | $0.82 \pm 0.12$ | $0.95 \pm 0.09$ | 0.795 | 0.938 |
| $50 < A \leq 100$ $(odd\ mass)$ | $0.85 \pm 0.09$ | $0.87 \pm 0.11$ | 0.704 | 0.681 |
| $100 < A \leq 150$ $(even\ mass)$ | $0.53 \pm 0.08$ | $0.92 \pm 0.06$ | 1.297 | 3.448 |
| $100 < A \leq 150$ $(odd\ mass)$ | $0.65 \pm 0.05$ | $1.05 \pm 0.07$ | 1.235 | 3.301 |

Table1. The ML-based estimated parameters of generalized Brody distribution in different sequences where $KLD_{GOE}$ and $KLD_{GUE}$ exhibit the distances of estimated distribution to GOE and GUE limits, respectively.

From Table1 and Figures (1,2), a deviation to non regular dynamic ($q \longrightarrow 1$),is apparent for these 6 sequence while inclination is dominant for odd-mass nuclei similar to suggested results by [31-32]. The KLD measures reveal similar results (we don't evaluate $KLD_{Poisson}$ for obvious chaoticity of sequences). These results prove the prediction of GOE where suggest spherical nuclei (magic or semi magic) exhibit shell model spectra and explore predominantly less regular dynamics in compare to deformed ones. Also, this result is known as AbulMagd-Weidenmuller chaoticity effect [33] where suggest the suppression of chaotic dynamics due to the rotation of nuclei. Also the less chaotic dynamics explore by heavier spherical nuclei in compare to light ones, reveal the prediction of Bae et al [27] while consider a chaos to order transition by spherical heavy nuclei. On the other hand, the KLD measures suggest a slight deviation to GUE statistics for spherical odd-mass nuclei in the $50 < A \leq 100$ mass region. As already mentioned in previous sections, due to the presence of discrete symmetries [34-35], one can expect GUE statistics of spectra although systems are time-reversal invariant (TRI). In this case, TRI induces degeneracy between different conjugate irreducible representations of the symmetry group (Kramers degeneracy). It means, such that TRI is broken within one of the irreducible subspaces of the point group. It was suggested that the eigenvalues belonging to these degenerate subspaces should have the statistical properties of the GUE rather than the GOE [36-37]. Since, Kramers degeneracy means the levels in an odd-fermions system are at least doubly degenerate while for spherical nuclei, (2j+1) degeneracy is reveal, these results can be considered as a significant identification of GUE-type statistics in nuclear systems. On the other hand, the deviation to GUE statistics in this sequence can't be regard as a predominant effect and for a overwhelming conclusion, one have to carry out the present statistics for all nuclei.

## 5. Summary and conclusion

We review the level statistics of spherical nuclei in the NNSD-based statistics by using all the available experimental data. With use of a generalized Brody distribution and also MLE technique, the chaoticity parameter is estimated with more precision. The difference in the chaoticity parameter of each sequence is statistically significant while the chaoticity parameter is dominant in the spherical odd-mass nuclei. Also, a trivial deviation to GUE statistics is apparent for odd-mass spherical nuclei in the $50 < A \leq 100$ where may be realized due to the irreducible subspaces of the point group. Our results may be interpreted a weaker pairing force between the single particle and collective degrees of freedom in odd-mass nuclei than even-mass ones.

## References

[1]. M. L. Mehta, Random Matrices, San Diego, Academic Press, 2nd ed(1991).

[2]. T. A. Brody, J.Flores, J.P.French, P.A.Mello, A.Pandey & S. S. M. Wong, Rev. Mod. Phys53(1981)385.

[3]. J. F. Shriner, G. E. Mitchell, and T. von Egidy, Z. Phys. A338 (1991)309.

[4]. A. Y. Abul-Magd and M. H. Simbel, J.Phys.G: Nucl. Part. Phys22(1996)1043-1052.

[5]. R. Schafer,M. Barth, F. Leyvraz, M. Mu¨ller, T. H. Seligman, and H.J. Stockmann, Phys. Rev.E66(2002) 016202.

[6]. A. Relaño, J. Retamosa, E. Faleiro, and J. M G. Gómez, Phys. Rev. E72(2005) 066219.

[7]. C. Dembowski, H.D. Graf, A. Heine, H. Rehfeld, A. Richter, and C. Schmit, Phys. Rev. E62(2000)R4516.

[8]. Gu Jing –Zhong, Wu Xi-Zhen, Zhuo Yi-zhong, Z. Phys. A354(1996)15.

[9]. F. Leyvraz, C. Schmit, and T. H. Seligman, J. Phys. A**29** (1996)L575.

[10]. F. Leyvraz and T. H. Seligman, in Proceedings of the IV, Wigner-Symposium in Guadalajara, Mexico, edited by N. M.Atakishiyev, T. H. Seligman, and K. B. Wolf World Scientific, Singapore(1996)p. 350.

[11]. J. F. Shriner, G. E. Mitchell, and Jr.Bilpuch, Z. Phys.A332(1989)45and Phys.Rev.Lett.59 (1987) 435, 59 (1987)1492.

[12]. A.Y. Abul-Magd et al, Phys. Lett. B579(2004)278-284.

[13]. T.von.Egidy, A.N.Behkami, H.H.Schmidt, Nucl.PhysA481(1988) 189and 454 (1986)109.

[14]. Declan Mulhall, Phys.Rev.C80(2009)034612 and 83(2011) 054321.

[15]. F. J. Dyson and M. L. Mehta, J. Math. Phys47(1963) 01.

[16]. A. Bohr and B. R. Mottelson , Nuclear structure, Vol II, Nuclear Deformation , Benjamin , New York, (1975).

[17]. W. Greiner and J. A. Maruhn, Nuclear Models, Springer-Verlag, Berlin Heidelberg, (1996).

[18]. A. Y. Abul-Magd and A.AL.Sayed, Phys. Rev. C74 (2006)037301.

[19]. A. Y. Abul-Magd and A.AL.Sayed et al, Nucl. Phys. A839 (2011)1.

[20]. A Al-Sayed, J. Stat. Mech. P02062 (2009).

[21]. M.A.Jafarizadeh , N.Fouladi,H.Sabri and B.R.Maleki submitted to Physics Scripta(nucl-th/1106.2497).

[22]. M.A.Jafarizadeh and et.al, submitted to J.Phys.G: Nucl. Part. Phys. (nucl-th/1101.0958).

[23]. National Nuclear Data Center,(Brookhaven National laboratory), chart of nuclides, (http://www.nndc.bnl.gov/chart/reColor.jsp?newColor=dm)

[24]. Live chart, Table of Nuclides, (http://www-nds.iaea.org/relnsd/vcharthtml/VChartHTML.html).

[25]. Richard B. Firestone,Virginia S. Shirley, S. Y. Frank , Coral M. Baglin and Jean Zipkin, table of isotopes, (1996).

[26]. P. Möller, J. R. Nix, W. D. Myers, W. J. Swiatecki, At. Data. Nucl. Data Tables59(1995)185.

[27]. M. S. Bae, T. Otsuka, T.Mizusaki and N.Fukunishi, Phys. Rev. Lett69(1992)2349.

[28]. T . A . Brody, Lett. Nuovo Cimento 7(1973)482.

[29]. A. Y. Abul-Magd et al, Ann. Phys. (N.Y.)321(2006)560-580.

[30]. M.V. Berry, M. Robnik, J. Phys. A: Math. Gen17(1984) 2413.

[31]. J. M. G. GÓMEZ-R. MOLINA-J. RETAMOSA, Published by World Scientific( 2002)pp. 273-281.

[32]. R.A. Molina, Eur. Phys. J. A28(2006)125-128.

[33]. Paar .V and Vorkapic.D, Phys.Lett.B, 205 (1988)7and Phys.Rev.C 41(1990)2397.

[34]. Xunjun Li and Jerzy Dudek, Phys. Rev. C49(1994)R1250.

[35]. Krzysztof Sacha, Jakub Zakrzewski and D.Delande, Phys. Rev. Lett83 (1999)2922.

[36]. J. B. French, V. K. B. Kota, A. Pandey and S. Tomosovic , Phys. Rev. Lett, 54(1985)2313.

[37]. S. Leoni, Journal of Physics: Conference Series 205(2010) 012041.

| Nuclei | N | $E_{max}$ | Nuclei | N | $E_{max}$ | Nuclei | N | $E_{max}$ | Nuclei | N | $E_{max}$ |
|---|---|---|---|---|---|---|---|---|---|---|---|
| $^{30}Si$ | 5 | 6537 | $^{30}P$ | 13 | 7605 | $^{30}S$ | 5 | 6766 | $^{31}S$ | 5 | 6996 |
| $^{32}Mg$ | 5 | 5203 | $^{32}Si$ | 5 | 5203 | $^{32}P$ | 15 | 8037 | $^{32}S$ | 6 | 7115 |
| $^{32}Cl$ | 6 | 3283 | $^{33}S$ | 10 | 5915 | $^{33}Cl$ | 7 | 5869 | $^{34}S$ | 10 | 6848 |
| $^{34}Ar$ | 6 | 7499 | $^{35}S$ | 13 | 7442 | $^{36}S$ | 5 | 7120 | $^{36}Cl$ | 5 | 3566 |
| $^{36}Ar$ | 12 | 7179 | $^{37}K$ | 12 | 6741 | $^{38}S$ | 6 | 5278 | $^{38}Ar$ | 6 | 5595 |
| $^{38}Ca$ | 6 | 5264 | $^{39}Ca$ | 5 | 6906 | $^{40}Ar$ | 6 | 4324 | $^{40}Ca$ | 29 | 9785 |
| $^{41}Ca$ | 7 | 5984 | $^{41}Sc$ | 16 | 6335 | $^{42}Ar$ | 6 | 5553 | $^{42}K$ | 5 | 3502 |
| $^{42}Ca$ | 11 | 4866 | $^{42}Ti$ | 5 | 4665 | $^{43}Ca$ | 10 | 6015 | $^{44}S$ | 5 | 3257 |
| $^{44}Ca$ | 11 | 4804 | $^{44}Ti$ | 23 | 9238 | $^{45}Ca$ | 10 | 5479 | $^{45}Sc$ | 5 | 3092 |
| $^{46}Ca$ | 12 | 7667 | $^{46}Ti$ | 23 | 6134 | $^{47}Ca$ | 10 | 9124 | $^{47}V$ | 5 | 6037 |
| $^{48}Ca$ | 10 | 8883 | $^{48}Ti$ | 10 | 4388 | $^{49}Ti$ | 7 | 6078 | $^{49}V$ | 5 | 5293 |
| $^{49}Cr$ | 7 | 6006 | $^{50}Ca$ | 5 | 4870 | $^{50}Ti$ | 6 | 4890 | $^{51}V$ | 5 | 5104 |
| $^{52}Ti$ | 6 | 4787 | $^{52}V$ | 6 | 2881 | $^{52}Cr$ | 15 | 5664 | $^{53}Cr$ | 10 | 7167 |
| $^{53}Fe$ | 8 | 7122 | $^{54}Mn$ | 24 | 4378 | $^{54}Fe$ | 14 | 6429 | $^{56}Fe$ | 28 | 5257 |
| $^{56}Ni$ | 8 | 8674 | $^{58}Ni$ | 26 | 6983 | $^{58}Co$ | 39 | 4082 | $^{59}Ni$ | 10 | 5957 |
| $^{114}Sn$ | 19 | 4030 | $^{116}Sn$ | 12 | 7179 | $^{118}Sn$ | 23 | 3944 | $^{120}Sn$ | 20 | 4190 |
| $^{122}Sn$ | 21 | 4284 | $^{124}Sn$ | 9 | 3265 | $^{126}Sn$ | 11 | 3964 | $^{128}Te$ | 16 | 3030 |
| $^{130}Te$ | 8 | 2745 | $^{131}Te$ | 7 | 3187 | $^{132}Te$ | 10 | 2918 | $^{132}Xe$ | 11 | 2959 |
| $^{136}Xe$ | 8 | 3212 | $^{136}Ba$ | 19 | 3706 | $^{138}Ba$ | 39 | 4665 | $^{138}Ce$ | 10 | 3356 |
| $^{139}Pr$ | 8 | 1532 | $^{140}Ba$ | 13 | 3527 | $^{140}Ce$ | 21 | 6187 | $^{140}Nd$ | 13 | 3561 |
| $^{142}Ce$ | 9 | 3697 | $^{142}Nd$ | 23 | 5355 | $^{142}Sm$ | 6 | 2747 | $^{144}Nd$ | 32 | 4845 |
| $^{145}Pm$ | 5 | 2112 | $^{145}Eu$ | 5 | 2494 | $^{146}Gd$ | 20 | 5228 | $^{146}Sm$ | 26 | 3693 |
| $^{148}Gd$ | 13 | 4051 | | | | | | | | | |

Table1. A description of available data where "N" describes the number of selected levels and $E_{max}$ represents the highest level contributed for each nucleus.

## Figure caption

**Figure1**. NNSD histograms displayed for spherical even-mass nucleus in different mass regions. Solid, dashed and dotted line represent GUE, Poisson and GOE curves respectively.

**Figure2.** Similar to Fig1, NNSD histograms displayed for spherical even-mass nucleus in different mass regions.

**Figure1.**

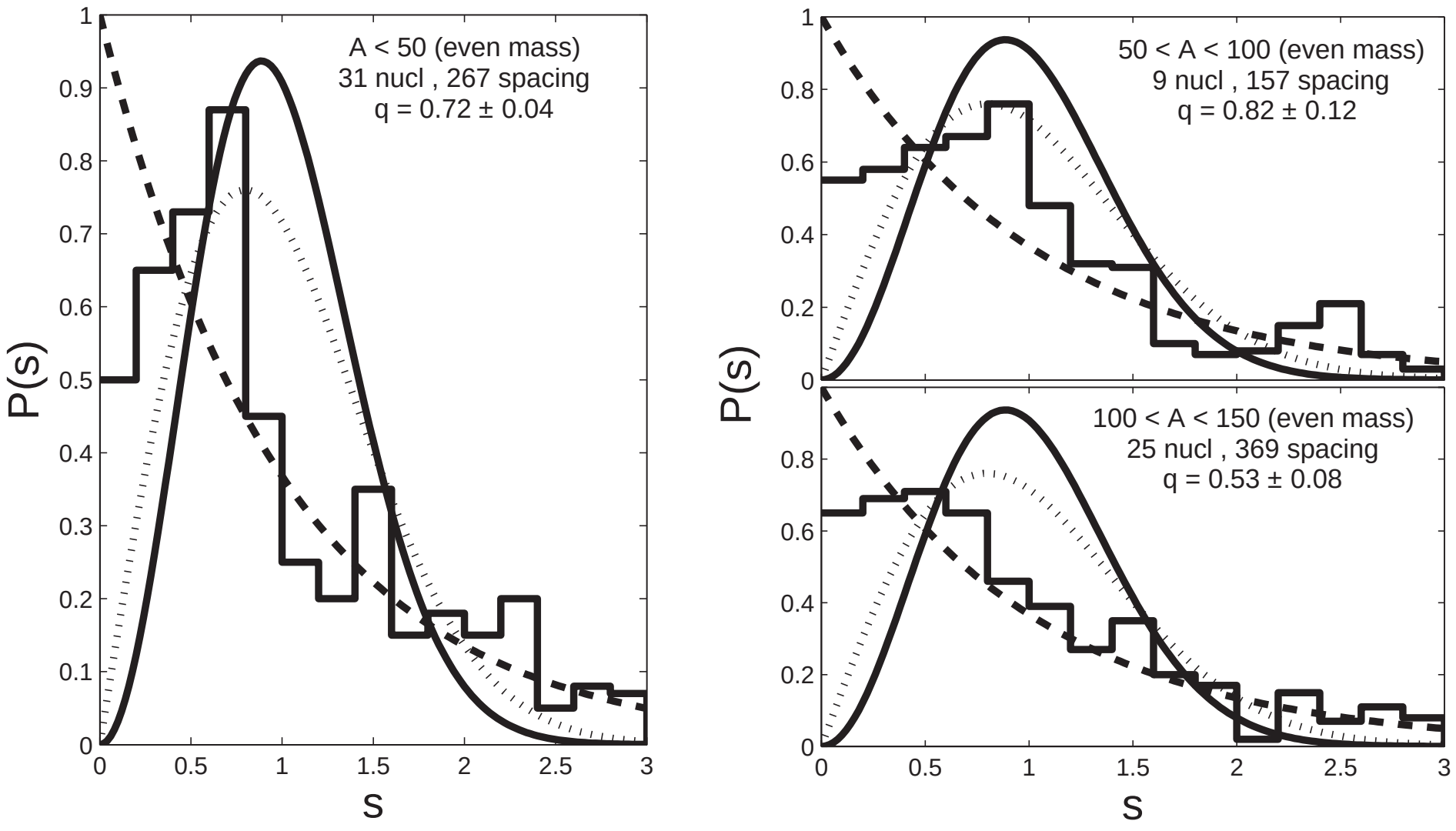

**Figure2.**

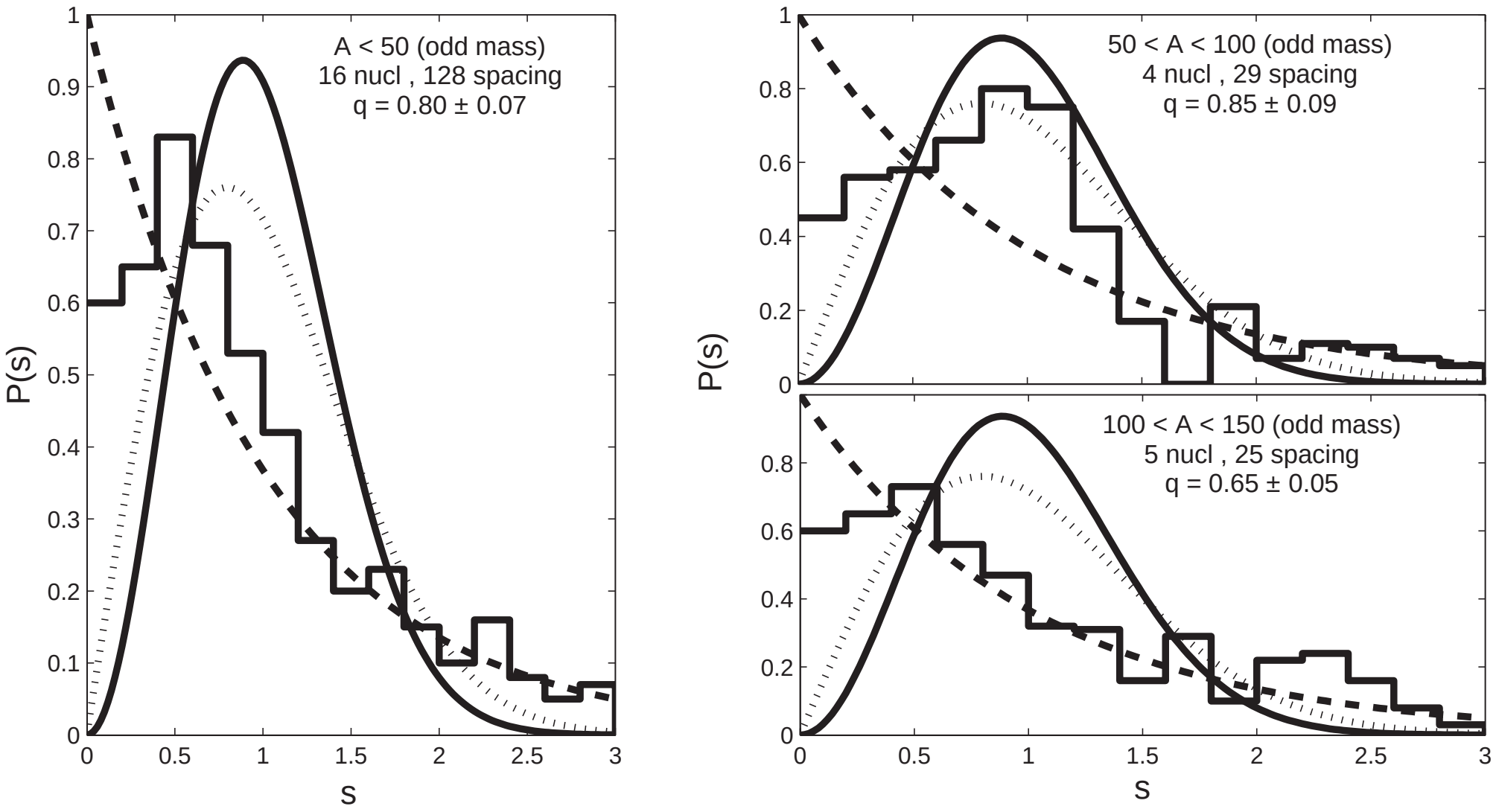